%% file: main.tex
\documentclass[sigconf]{acmart} 
\pdfoutput=1

\AtBeginDocument{%
  \providecommand\BibTeX{{%
    \normalfont B\kern-0.5em{\scshape i\kern-0.25em b}\kern-0.8em\TeX}}}

\copyrightyear{2022}
\acmYear{2022}
\setcopyright{acmcopyright}\acmConference[WebSci '22]{14th ACM Web Science Conference 2022}{June 26--29, 2022}{Barcelona, Spain}
\acmBooktitle{14th ACM Web Science Conference 2022 (WebSci '22), June 26--29, 2022, Barcelona, Spain}
\acmPrice{15.00}
\acmDOI{10.1145/3501247.3531565}
\acmISBN{978-1-4503-9191-7/22/06}

\usepackage{array, multirow}
\newcolumntype{L}[1]{>{\arraybackslash}p{#1}}
\usepackage{subcaption}
\usepackage{tikz}
\newcommand*\circled[1]{\tikz[baseline=(char.base)]{
            \node[shape=circle,draw,inner sep=.5pt] (char) {#1};}}
\usepackage[linesnumbered,algoruled,boxed,lined]{algorithm2e}
\DontPrintSemicolon

\usepackage[shortlabels]{enumitem}
\usepackage{siunitx}

\begin{document}

\title[It's the Same Old Story! Enriching Event-Centric Knowledge Graphs by Narrative Aspects]{It's the Same Old Story! Enriching Event-Centric Knowledge Graphs by Narrative Aspects}

\author{Florian Plötzky}
\email{ploetzky@ifis.cs.tu-bs.de}
\orcid{0000-0002-4112-3192}
\affiliation{%
  \institution{Institute for Information Systems, TU Braunschweig}
  \streetaddress{Mühlenpfordtstr. 23}
  \city{Braunschweig}
  \state{Lower Saxony}
  \country{Germany}
  \postcode{38106}
}  

\author{Wolf-Tilo Balke}
\email{balke@ifis.cs.tu-bs.de}
\orcid{0000-0002-5443-1215}
\affiliation{%
  \institution{Institute for Information Systems, TU Braunschweig}
  \streetaddress{Mühlenpfordtstr. 23}
  \city{Braunschweig}
  \state{Lower Saxony}
  \country{Germany}
  \postcode{38106}
}   

\pdfstringdefDisableCommands{%
  \def\\{}%
}


\renewcommand{\shortauthors}{Florian Plötzky and Wolf-Tilo Balke}

\begin{abstract}
Our lives are ruled by events of varying importance ranging from simple everyday occurrences to incidents of societal dimension. 
And a lot of effort is taken to exchange information and discuss about such events: generally speaking, \emph{stringent narratives} are formed to reduce complexity.
But when considering complex events like the current conflict between Russia and Ukraine it is easy to see that those events cannot be grasped by objective facts alone, like the start of the conflict or respective troop sizes. 
There are different viewpoints and assessments to consider, a different understanding of the roles taken by individual participants, etc. 
So how can such subjective and viewpoint-dependent information be effectively represented together with all objective information?
Recently \emph{event-centric knowledge graphs} have been proposed for objective event representation in the otherwise primarily entity-centric domain of knowledge graphs.
In this paper we introduce a novel and lightweight structure for event-centric knowledge graphs, which for the first time allows for queries incorporating viewpoint-dependent and narrative aspects. 
Our experiments prove the effective incorporation of subjective attributions for event participants and show the benefits of specifically tailored indexes for narrative query processing.
\end{abstract}


\maketitle

\section{Introduction}
\label{sec:introduction}
\input{sections/01-Introduction}

\section{Formalizing Events}
\label{sec:formalization}
\input{sections/02-NarrativePrototypes}

\section{Event Representation and Prototype Evaluation}
\label{sec:representation}
\input{sections/03-Representation}

\section{Proof of Concept Evaluation}
\label{sec:poc}
\input{sections/04-Evaluation}

\section{Related Work} 
\label{sec:related_work}
\input{sections/05-RelatedWork}

\section{Summary \& Outlook}
\label{sec:conclusion}
\input{sections/06-Conclusion}

\begin{acks}
This work was supported by the Leibniz-ScienceCampus Postdigital Participation operated by the Leibniz Association (Leibniz-Ge\-mein\-schaft). The first author likes to thank all raters for their work in the witness assessment evaluation.
\end{acks}
\bibliographystyle{ACM-Reference-Format}
\bibliography{biblio}
\end{document}

%% file: sections/01-Introduction.tex
The Web as today's prime resource of knowledge has drastically changed its structure over time.
It evolved from a rather unstructured Web of documents into a Web of structured data.
Indeed, since the first draft of a Semantic Web, technologies for information extraction, linked open data sources, and knowledge graphs (KGs) set the standard for structured knowledge representation on the Web and thus served as a key enabler for semantically richer applications.

Taking a closer look at large structured knowledge sources, such as DBPedia~\cite{auer2007dbpedia}, Wikidata~\cite{vrandevcic2014wikidata}, or YAGO~\cite{suchanek2007yago}, reveals that most information represents factual knowledge about real world entities.
For instance, the birth date of some person, the population count of a city, or the web page of some organization state factual and ready to use entity-centric information.
In contrast, the \emph{representation of events} is more difficult beyond time and location.
Although event-centric information is increasingly important, it generally offers a richer structure: events consist of several entities engaging in an (often complex) interaction bound by time and location.
While the obvious solution of representing simple facts on events as part of  a knowledge graph may seem enticing, see e.g., \cite{gottschalk2018eventkg, gottschalk2021oekg}, it falls short of actually solving the problem.
The vast collections of news articles, political commentary, and user reviews still offered on the Web today may serve to corroborate this point.

\begin{table*} 
    \centering
    \caption{Examples for different attribution predicates regarding a conflict situation.}
    \label{tab:ex_attributions}
    \begin{tabular}{@{} lL{0.25\textwidth}L{0.25\textwidth}L{0.25\textwidth} @{}}
        \toprule
        & \emph{Entity attributions} & \emph{Event attributions} & \emph{Attrib. for entities in events} \\
        \midrule
        \emph{objective} & possesses\_nuclear\_weapons & happened\_during\_cold\_war & is\_underdog\\
        \emph{subjective} & \centering - & is\_potential\_war\_starter & is\_aggressor \\
        \bottomrule
    \end{tabular}
\end{table*}

The reason is that when dealing with event-centric information, humans assume an \emph{intrinsic narrative structure} to make sense of complex events \cite{laszlo2008scienceofstories}. 
Such a narration can be understood as a textual or graph-based description of the sequential nature and individual salient steps involved in the interaction between the participants within some event.
Depending on these narratives' plausibility, information about events can be efficiently exchanged between humans and put into perspective.
Consider for example the tensions between Russia and Ukraine in the advent of the Russian invasion of Ukraine on February 24, 2022.\footnote{Please note that all examples in this paper regarding Russia and the Ukraine are related to the conflict between both nations \textbf{before} the invasion started.} 
Factual event-centric information regarding these tensions are the location and time along with the participants and the point of contention.
However, for actually understanding the conflict a more discourse-oriented position is needed: apart from the factual correctness or trustworthiness of information about the conflict, a variety of differing opinions, viewpoints, and sentiments have to be taken into account. 
For example, the tensions started after the takeover of the Crimean Peninsula by Russian troops which was coined as an \emph{"illegal annexation"} by the Ukraine and as a \emph{"legal secession following a referendum"} by the Russian side, cf. \cite{mamlyuk2015ukraine}.
Moreover, narratives may strongly differ also in other aspects like the participants (e.g., regarding the role of the European Union) or the involvement of other organizations such as the NATO.

In brief, facts about events may be \emph{multi-faceted}, \emph{inconsistent}, and \emph{subjective}.
Of course, Semantic Web technologies already do allow for a basic representation of facts covering different, even contrasting aspects.
However, while techniques like reification and quantifiers are commonly used today, they add a higher level of complexity, cause problems to derive consistent information and severely hamper retrieval efficiency (see \cite{rouces2015framebase} for a good explanation on  problems of consistency and complexity and \cite{hernandez2015reifying} for a detailed performance analysis of different reification implementations). 

In a recently published paper \cite{ploetzky2021analogies} we introduced the vision of \emph{narrative prototypes} as a means to capture narrative aspects for events.
However, the narrative prototypes have only been roughly sketched  without any formal notation or further details on how to actually implement them. 
In this paper we give a detailed elaboration of narrative prototypes and develop a formal representation to utilize them for querying for events in an innovative manner.
We propose a light-weight schema for enriching event-centric KGs by narrative information to enable them for event-centric narrative queries. 
Furthermore, we demonstrate the capabilities of the enriched KGs in a proof of concept built on a novel retrieval process in tight combination with textual knowledge.
Finally, we show how the usage of specially-tailored indexes increase the query effectiveness by a factor of 3 for our proof of concept.

%% file: sections/02-NarrativePrototypes.tex
In the following section we first provide a formal definition of events, event types, and participants.
Afterwards, subjective and objective attributions are introduced as a means to further characterize events.
Finally, we formally introduce narrative prototypes based on our previous definitions.

\subsection{On Events, Types, and Participants}
\label{subsec:on_events}
In general, we describe events as interactions between participants that happen at a given place to a known time.
Therefore, events are instantiated in a certain time period, i.e., a point in time or a time interval, at a specific location. 
We denote time as $T = T_I \overset{\cdot}{\cup} T_P$, where $T_I$ denotes the set of time intervals and $T_P$ the set of points in time.
Furthermore we define $L$ to be the set of event locations and $\Sigma_l$ as the set of event labels.
Events $E$ are denoted as the Cartesian product of time, locations, and event labels, i.e.:
\begin{equation}
    \label{eq:events}
    E = T \times L \times \Sigma_l
\end{equation}

Event labels $\sigma_l \in \Sigma_l$ should be unique identifiers for their respective event.
Beside the label, events are instances to at least one event type $t \in \mathcal{T}$.
We define a function \emph{event\_types} which maps each event $ev$ to a set of event types:
\begin{equation}
    \label{eq:event_types}
    \textit{event\_types}: \, ev \mapsto \{t_1, \dots, t_n\},\,\,t_i \in \mathcal{T}, ev\in E
\end{equation}

Events can therefore be instances of one or multiple event types. 
Furthermore, $\mathcal{T}$ can be modeled in a hierarchical structure, i.e., as taxonomy or thesaurus.
Beside this structure we argue, that, due to different  granularities of $T$, $L$ and  $\mathcal{T}$, modeling the codomain of \emph{event\_types} as set is more suitable than the assignment of a single type.
The main reasons are highly entangled events, e.g., fraud elections leading directly to riots on the same day, or disputable or unknown events.

Additionally, in theory $L$ and $T$ could be coarse grained and encompass large areas of space and time, i.e., events could encompass decades as a time interval and the whole world as location.
In this paper we limit the scope to events with a short time interval and a small region.
Valid events are for example the Gulf and Iraq Wars in 1991 and 2003 respectively or the Vietnam war between North Vietnam and the USA.
We exclude event-like concepts like the "Cold War" or the "War on Terror".

Each event can involve a number of participants.
A participant is an entity $e \in \mathcal{E}$ which is defined as a real-world object that can be further described by a number of properties.
Analogous to the event types we define a \emph{participants} function to assign participants to events:
\begin{equation}
    \label{eq:participants}
    \textit{participants}: \, ev \mapsto \{e_1, \dots, e_m\}, \,\, e_i \in \mathcal{E}, ev \in E
\end{equation}

Each participant encompasses a set of attributes $A_e \subset \mathcal{A}$ which may be further qualified by time.
Additionally, participants in events are further characterized by \emph{event roles} $\Sigma_r$.
Event roles are used to describe the relation between the event and a participant, e.g., we can use the event role \emph{winner} to express that a certain participant won in a event of the type \emph{conflict}.
We define two functions: 
\begin{equation}
    \label{eq:roles}
    \textit{roles} :\, t \mapsto \{\sigma_1, \dots, \sigma_r\},\,\, \sigma_i \in \Sigma_r, t \in \mathcal{T}
\end{equation}
\begin{equation}
    \label{eq:role_participant}
    \text{\emph{role}} : E \times \mathcal{E} \rightarrow \Sigma_r
\end{equation}

Eq.~\ref{eq:roles} assigns the set of possible roles for an arbitrary event type.
Since events can have multiple event types, the set of possible roles for an event $ev$ is defined as $\bigcup_{t \in \text{\emph{event\_types}}(ev)} \text{roles}(t)$.
Eq.~\ref{eq:role_participant} is used to assign a role to a particular participant in an event.


\subsection{Attributions}
\label{subsec:attributions}

\begin{figure*}
    \centering
    \begin{subfigure}{0.5\linewidth}
        \centering
        \includegraphics{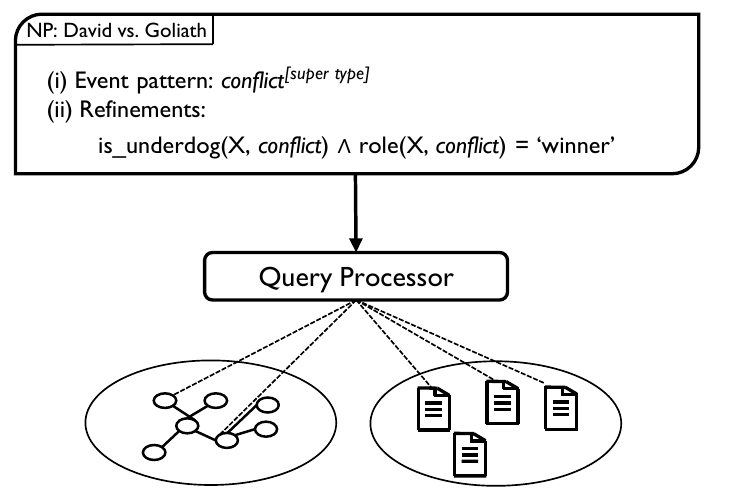}
        \caption{The David vs. Goliath prototype.}
        \label{subfig:npt_ex1}
    \end{subfigure}\hfill
    \begin{subfigure}{0.5\linewidth}
        \centering
        \includegraphics{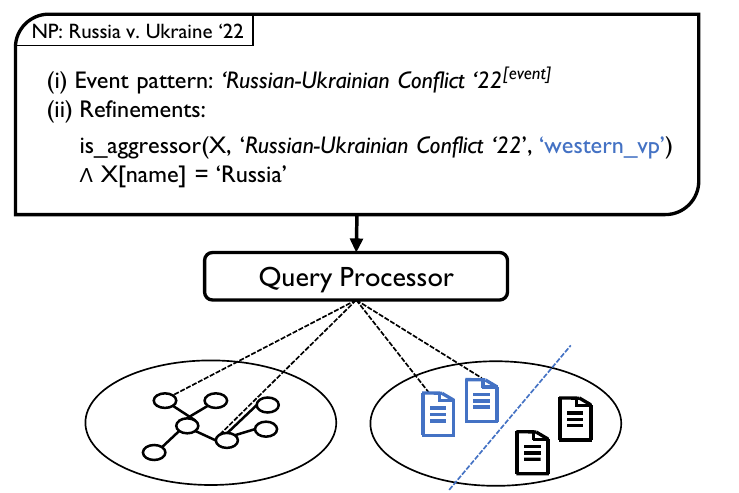}
        \caption{Asking for an aggressor from a western viewpoint in the Russian-Ukrainian conflict.}
        \label{subfig:npt_ex2}
    \end{subfigure}
    \caption{Examples for Narrative Prototypes.}
    \label{fig:narrative_prototype_example}
\end{figure*}

Beside the pure structural description of events, i.e., the type, time, location, and participants, news articles, social media posts, or other textual descriptions regarding events oftentimes make certain attributions to an event or its participants.
The before mentioned Russian-Ukrainian conflict and the role of Russia or the Ukraine respectively can be seen as an example.
Publications regarding the conflict are likely to frame certain parties of the conflict creating a hegemonic frame \cite{macgilchrist2011journalism} which we call a \emph{viewpoint} in this paper.
We denote the set of viewpoints as $\mathcal{V}$.
Analogous to $L$ and $T$, $\mathcal{V}$ can have different granularities, i.e., viewpoints can be defined for organizations, persons, or even abstract entities like the "western world".

Attributions are encoded as predicates, e.g. 
\begin{equation*}
  \text{is\_underdog} : \mathcal{E} \times E \rightarrow \{0, 1\}
\end{equation*}
which states whether the participant $e$ can be seen as an underdog in event $ev$.
We define attributions for events, entities, and entities as event participants, where the is\_underdog predicate is an example for the latter.

Additionally, attributions can be subject to a certain viewpoint. This is the case for instance for the \emph{aggressor} attribution in the Russian-Ukrainian conflict where it depends on the source material whether the claim that Russia is an aggressor holds or not.\footnote{cf.  \cite{wong2022natoukraine} and \cite{amar2022natoukraine} as an example for opposing viewpoints on the Russian-Ukrainian conflict}
If a viewpoint is necessary for the evaluation of an attribution we call it \emph{subjective attribution} otherwise it is called an \emph{objective attribution}.
As an example, the subjective aggressor predicate can be defined as:
\begin{equation*}
  \text{is\_aggressor} : \mathcal{E} \times E \times \mathcal{V} \rightarrow \{0, 1\}
\end{equation*}

Table~\ref{tab:ex_attributions} provides some examples for subjective and objective attributions for entities, events and entities as participants in events regarding the current Russian-Ukrainian conflict. 
Note that since we focus on event-centric repositories in this paper, subjective attributions for entities are excluded.
Therefore such attributions can only be made in the context of events.

\subsection{Narrative Prototypes}
\label{subsec:narrative_prototypes}
After we introduced a formal description of events and defined objective and subjective attributions, we can now formally introduce \emph{narrative prototypes}.
Verbally, a narrative prototype describes an abstract event pattern along with attributions regarding the pattern.
The main idea here is to express complex events in terms of their structural properties (instance of the given set of event types) and their perception (i.e., query refinements utilizing attributions regarding the event).
Therefore we define a narrative prototype as a template for events consisting of two components:
\begin{description}
    \item[Event Pattern] This component can either be an event, an event type, or an event super type. 
    The latter requires $\mathcal{T}$ to be modeled in a hierarchical structure, i.e., as taxonomy.
    If the event pattern is an event type $t \in \mathcal{T}$, any event $ev \in E$ with $t \in \text{\emph{event\_type}}(ev)$ is matched by the pattern.
    If the event pattern is an event super type $st \in \mathcal{T}$ any event $ev \in E$ can be matched by the pattern if either the super type or any sub type of the super type is in $\text{\emph{event\_type}}(ev)$.
    \item[Refinements] The second component is a set of refinements for the event pattern. 
    A refinement is a logical expression consisting of subjective and objective attributions and event functions as introduced in Sec.~\ref{subsec:on_events}.
    If qualified, attributions to entities in narrative prototypes are limited to the time component of the event matched by the pattern. For example if the Ukrainian-Russia conflict in 2022 is matched by the event pattern attributions like the before mentioned possesses\_nuclear\_weapons for Russia are evaluated for the year the conflict takes place.
    If event functions like \emph{event\_types} are used, they must be part of an equation to yield a Boolean value.
\end{description}

In a narrative analogy the event pattern defines the basic story block (a conflict, a festival, an election, etc.) and the refinements are narrative elements to further characterize the event and its participants.
Events that are \emph{matched} by the event pattern and for which all attributions are true are considered as a match for the narrative prototype.
Note that the second requirement could be weakened to a certain number of attributions which must be true for a match.

Narrative prototypes can be used to query events from event-centric repositories in an expressive way. 
In contrast to keyword queries the prototypes preserve the context of the attributions and should therefore increase the precision of the query (but will most likely lower the recall).
Another example for the usage of event-centric narrative queries is the verification of a specific set of assumptions regarding a particular event.
In this case, the supertype is replaced by a concrete event and only the attributions are evaluated.
During this process the subjective attributions are evaluated regarding their respective viewpoints.
If no viewpoint is given the subjective attribution holds if the attribution is true in at least one viewpoint.
Fig.~\ref{fig:narrative_prototype_example} depicts the usage of both variants for an event repository consisting of an event-centric knowledge graph and a document collection.

In Fig.~\ref{subfig:npt_ex1} a query by prototype is depicted as a David vs. Goliath situation.
First, all events are considered whose types include at least one conflict type, i.e., the event type \emph{conflict} or any subtype of \emph{conflict}.
For each of those events, the refinements are applied.
In this example we check for the participant in role \emph{winner} whether this entity can be seen as an underdog in the conflict or not.
Example matches of this prototype are the Vietnam war in the late 1960s where the USA lost the war against North Vietnam or the Battle of Brownstone in the USA where an army of 200 US soldiers lost against two dozen native Americans.

The second example in Fig.~\ref{subfig:npt_ex2} depicts a prototype asking whether an entity with the name "Russia" was an aggressor in the Russian-Ukrainian Conflict in 2022 (RUC22) from a western viewpoint.\footnote{We assume that $\textit{name} \in A_X$}
The western viewpoint serves as a filter in the query processing and limits the query scope to sources with a western viewpoint as indicated by the blue cutting line.

%% file: sections/03-Representation.tex
In this section we discuss and provide a suitable and lightweight representation for events and narrative prototypes.
Furthermore, we discuss how narrative prototypes can be evaluated concerning subjective and objective attributions.

\subsection{Event Representation}
\label{subsec:ev_representation_kgs}
Evaluating narrative prototypes requires a representation model to encompass events $E$, event types $\mathcal{T}$, entities $\mathcal{E}$ as participants, and viewpoints $\mathcal{V}$.
Additionally, we require the model to allow for evaluation of objective and subjective attributions, i.e., event roles and viewpoint-dependent data must be accessible in an efficient manner.
Over the years, various models for representing events based on the Resource Description Framework (RDF) or upper layer ontologies have been proposed, e.g., LODE~\cite{shaw2009lode}, Event-Model-F~\cite{scherp2009feventmodel} or the Simple Event Model (SEM)~\cite{vanhage2011sem}.
All three of them can represent events, participants, and event types and feature a rich set of predicates to connect events and participants, e.g., sub-events or types for participants.
However, viewpoints on events as a key feature for subjective attributions (see Sec.~\ref{subsec:narrative_prototypes}), are not supported by most models except for SEM.
\begin{table*}
    \centering
    \begin{tabular}{lrrrrrr}
    \toprule
    &Crimea Crisis & Gulf War & Iraq War & Ukraine Conflict & Vietnam War & $\Sigma$ \\
    \midrule
    New York Times & 79 & 129 & 129 & 79 & 42 & 458\\
    Washington Post & 79 & 83 & 106 & 79 & 54 & 401\\
    \midrule
    Daily Mail & 91 & 28 & 70 & 66 & 12 & 267\\
    The Guardian & 78 & 51 & 127 & 79 & 1 & 336\\
    \midrule
    RT.com & 78 & 10 & 73 & 78 & 5 & 244\\
    Sputniknews & 79 & 8 & 10 & 79 & 1 & 177\\
    \bottomrule
    \end{tabular}
    \caption{Corpus statistics: number of articles per outlet for the selected conflicts}
    \label{tab:corpus_statistics}
\end{table*}
Viewpoints in SEM are represented by using RDF reification. 
Each participant can have a role in an event that varies depending on the viewpoint.
An authority in the representation can be used to define which viewpoint is used to derive a participants' role.
The original work on SEM \cite{vanhage2011sem} depicts a police action event in Indonesia by police forces of the Netherlands as an example. For this event, the Netherlands are depicted as occupiers or liberators depending on whether the authority is Indonesia or the Netherlands.

In theory, we could use SEM roles to represent subjective attributions for each participant in each event, build a triple store based on, it and utilize standard query languages like SPARQL to evaluate objective and subjective attributions.
However, reification leads to a high amount of triples \cite{rouces2015framebase}, especially if $|\mathcal{V}|$ is large.
SEM would require four additional triples for each viewpoint on an event role, i.e., four triples for each possible interpretation of a subjective attribute.
Additionally, the complexity of SPARQL queries would increase.

\begin{figure}
    \centering
    \includegraphics{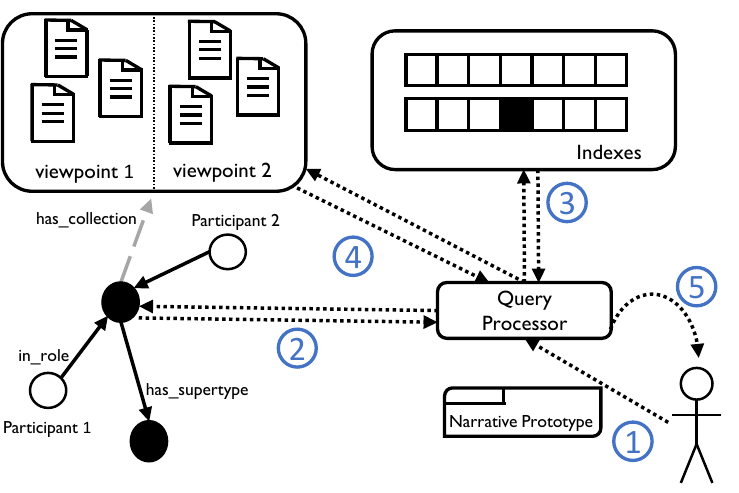}
    \caption{System Sketch}
    \label{fig:system_sketch}
\end{figure}

Knowledge graphs built on SEM like EventKG do not store view\-point-related information as of today.
Information regarding viewpoints is mostly stored in texts regarding events.
Most event extraction techniques, however, focus on event structures and do not explicitly capture different views regarding the events \cite{xiang2019eventsurvey}.
We argue that this dichotomy between the representation of structural information (e.g., participants and sub-events) in knowledge graphs and viewpoint-dependent information in texts is natural and should not be forced into a single representation.
Recently published research regarding the semantic enrichment of knowledge graph with news articles follow this approach \cite{rudnik2019searching, rospocher2016kgnews}.

For this paper, any event-centric knowledge graph with support for event types and event roles for participants is generally suitable.
As a minimum requirement, we define the knowledge graph $KG(E\cup \mathcal{E}\cup \mathcal{T}, A)$ as a directed graph.
The nodes in the graph represent events, event types, and entities. 
Directed edges are represented by $A$ and connect entities, events, and event types either by event functions (as discussed in Sec.~\ref{subsec:on_events}), super-type relationships to model $\mathcal{T}$ as taxonomy, or objective attributions like is\_underdog.
Furthermore we denote a special edge $\text{has\_collection} \in A$ which links events to a document collection.
The collection should only include documents regarding the particular event and each document is assigned to at least one viewpoint.
The left portion of Fig.~\ref{fig:system_sketch} depicts a possible $KG$ consisting of an event with two participants, an event type, and a link to a document collection consisting of two viewpoints.

\subsection{Narrative Prototype Evaluation}
\label{subsec:np_evaluation}
Narrative prototypes are designed by a user as a means to query events in an expressive way from an enriched event-centric KG.
As described in Sec.~\ref{sec:formalization}, the prototypes consist of event patterns and refinements in form of objective and subjective attributions regarding the event pattern.
A prototype matches an event if the event fulfills both, the event pattern and all refinements.
Fig.~\ref{fig:system_sketch} depicts this process in five steps which are explained in detail in the following.

Users first have to design a narrative prototype and pass it to the query processor \circled{1}, i.e., they define an event pattern and refine it by using attributions.
Examples for narrative prototypes are given in Fig.~\ref{fig:narrative_prototype_example}. 
The system may give suggestions on which attributions are possible for a given event pattern based on a query log. 
While the event pattern and the objective attributions can be evaluated by using well-known query languages, the subjective attributions require iterations over a set of documents and should be reduced beforehand if possible. 

Step  \circled{2} in the system involves querying the knowledge graph $KG$ to: 1. decide which events match the event pattern and 2. the evaluation of the objective attributions.
If the event pattern is an event type or supertype $t \in \mathcal{T}$ all events with $t \in event\_types(t)$ are candidates to match the prototype, otherwise the given event is the only candidate.
For each candidate we have to check whether the objective attributions hold.
Objective attributions can consider:
\begin{itemize}
    \item The structure of the event itself, e.g., time, location or number of participants.
    \item Attributes of the participants involved in the event, e.g., the name of a country or the requirement that at least one participant possesses nuclear weapons.
    \item Participants in the event, i.e., the event roles a participant is assigned to regarding the event (cf. Eq.~\ref{eq:role_participant}).
    Event roles can be rather general, e.g., winners, referees, or commentators, or already exhibit a rather narrative character like \emph{underdog}.
    In distinction to subjective attributions all objective attributions must not include a viewpoint and it must be possible to evaluate them unambiguously.
\end{itemize}
In practical terms it should be possible to evaluate both, the event pattern as well as objective attributions with suitable query languages like SPARQL for triple stores or Gremlin for graph databases if $KG$ fulfills the requirements as discussed in Sec.~\ref{subsec:ev_representation_kgs}.
Events not matching the event pattern or missing at least one objective attribution are not considered for the next steps in the evaluation processes.
For the remaining events we collect the pointers to the document collections for each respective event.

Regarding the evaluation of subjective attributions, we first rely on a specially tailored index structure in \circled{3}.
For each possible subjective attribution $s$ and for each participant $e \in \mathcal{E}$ we generate an index regarding whether $s(e)$ was true for any event in any viewpoint.
It is also possible to introduce a finer grained index structure that includes all viewpoints $v \in \mathcal{V}$ and checks whether $s(e, v)$ holds for any event.
In both cases the indexes act as a filter in two cases: 1. if a prototype specifically requires a certain participant to be in a subjective attribution $s$ (e.g., the prototype in Fig.~\ref{subfig:npt_ex2}) we can stop the evaluation if $s$ is never evaluated to be true for the participant and 2. we can reduce the number of participants to check during the evaluation to those for which $s$ holds.

All remaining events are then passed to the document-based evaluation in \circled{4}.
Here, the document collection for each event is further narrowed down to the viewpoints required in the subjective attributions.
For all remaining documents, Information Retrieval (IR) or Natural Language Processing (NLP) techniques are applied to verify whether a document supports the subjective attribution or not.
We call a document supporting a subjective attribution a \emph{witness} w.r.t. the attribution.
Witnesses can further be ranked based on certain features, e.g., one might value opinion articles from well-known domain experts more than a news ticker article.
If the number of witnesses surpasses a certain threshold the respective subjective attribution is evaluated as true.
Again, all events for which at least one subjective attribution does not hold are eliminated.
Note that this step is costly since IR or NLP techniques are involved.
For this reason the event pattern and objective attributions are evaluated first in step \circled{2} to reduce the number of necessary subjective evaluations.
The index structure \circled{3} and viewpoint restrictions further shrink the number of participants and potential witnesses to be evaluated.
Thus, subjective attributions can be evaluated efficiently.
Finally in step \circled{5} the remaining events are returned to the user as matches to her narrative prototype.

%% file: sections/04-Evaluation.tex
Based on Fig.~\ref{fig:system_sketch} we developed a proof of concept query processor based on a small number of events, viewpoints, and attributions to test the query expressiveness and efficiency of our proposed system.
Note that this proof of concept is limited in scale and is intended to provide a first attempt to tackle the problem of evaluating narrative prototypes and reveal limitations and future work items for this task.
Furthermore, since the evaluation of event patterns and objective attributions can mostly be done by well-known graph query languages, we focus on the subjective attributions, i.e., steps \circled{3} and \circled{4} in Fig.~\ref{fig:system_sketch}.

We formulate two research questions:
\begin{description}
    \item [\textbf{RQ1}] How can we assess witnesses for subjective attributions by using state-of-the-art NLP techniques and what are problems and open questions for this use case?
    \item [\textbf{RQ2}] How can we design suitable index structures for subjective attributions and what are the potential benefits?
\end{description}
To answer the questions we constructed a small knowledge graph based on the requirements in Sec.~\ref{subsec:ev_representation_kgs} based on a set of events.
Additionally, we collected several news articles for each event from different outlets to simulate different viewpoints.
The experimental setup is discussed in detail in Sec.~\ref{subsec:doc_collection}. 
The two research questions are addressed in Sections \ref{subsec:witness_assessment} and \ref{subsec:indexing}, respectively.

\subsection{Event-Centric KG and Document Collection}
\label{subsec:doc_collection}
For this proof of concept we selected the following conflict situations (wars and international tensions):
\begin{enumerate}
    \item the second Indochina War (Vietnam War between North and South Vietnam and their respective allies),
    \item the second Gulf War between Iraq and Kuwait in 1990/91,
    \item the Iraq War in 2003,
    \item the Crimea crisis leading to the accession of Crimea to Russia in 2014 and
    \item the current Russian-Ukrainian conflict (i.e., the conflict before the war started).
\end{enumerate}
We constructed a small knowledge graph for each event, the participants, and some objective attributions, e.g., North Vietnam and Iraq are underdogs in the Vietnam war and Iraq war, respectively.
Furthermore, we defined three coarse viewpoints based on certain countries, i.e., United States (US), United Kingdom (UK), and Russia (RU).
We selected two well known news outlets for each viewpoint: \emph{The New York Times} and  \emph{The Washington Post} for US, the \emph{Daily Mail} and \emph{The Guardian} for UK, and \emph{Russia Today} (RT.com) and \emph{Sputniknews} for RU.
Please note: we do \textbf{not} state that those media outlets represent a homogeneous view or can be seen in any way as representatives for the viewpoints of any respective countries' administrations or citizens.
To find suitable articles we formulated a Boolean search query for (1)--(5) in DuckDuckGo\footnote{\url{https://duckduckgo.com/}} (DDG) and used the \emph{site} qualifier to specifically search on the websites of the mentioned outlets.
The reason for utilizing DDG instead of specialized platforms like MediaCloud~\cite{roberts2021mediacloud} was the mostly missing RU perspective.
Hence, we relied a general search engine to build the document collection.
During a quality check, obvious mistakes like articles featuring topics like vaccinations in Ukraine have been filtered out.
Afterwards, we collected the headlines and news text of each link found at DDG if available in our subscription.
Tab.~\ref{tab:corpus_statistics} depicts the number of articles for each outlet and event.
The numbers are still skewed especially for the Vietnam, Gulf and Iraq War because: firstly we only collected recent articles and the Vietnam and Gulf war are events of the past and secondly Russia was not directly involved in those events (except for diplomatic reasons) making them less relevant for RT.com and Sputniknews.
Nevertheless, the size of the respective corpora is not relevant for \textbf{RQ1} and \textbf{RQ2}.

\subsection{Witnesses Assessment}
\label{subsec:witness_assessment}
After constructing a small document corpus and knowledge graph we can now proceed to the first research question regarding the assessment of witnesses. In the context of answering \textbf{RQ1} we define a set of subjective attributions concerning conflict situations. In particular we constructed four subjective attributions:
\begin{enumerate}[(a)]
    \item  $\text{is\_aggressor} : \mathcal{E} \times E \times \mathcal{V} \rightarrow \{0,1\}$ 
    \item  $\text{is\_threat} : \mathcal{E} \times E \times \mathcal{V} \rightarrow \{0,1\}$ 
    \item  $\text{is\_enemy} : \mathcal{E} \times E \times \mathcal{V} \rightarrow \{0,1\}$ 
    \item  $\text{is\_war\_criminal} : \mathcal{E} \times E \times \mathcal{V} \rightarrow \{0,1\}$ 
\end{enumerate}

Attribution (a) asserts "aggressive behavior" to a participant in a certain event.
The exact meaning of "aggressive" is subjective information and may depend on the witnesses of the respective viewpoints.
Similarly the definition of "threat" for attribution (b) depends on the witnesses.
Note that (a) and (b) overlap up to a certain degree since aggressors are usually threats to others.
However, a participant of an event can still be a threat without being an aggressor, e.g.,  in proxy wars the nation behind the proxy is often attributed to be a threat without being aggressive.
Hence a proxy war and no direct warfare.
While (a) and (b) can mostly be attributed from any viewpoint for any event, (c) is more likely to be used in viewpoints of the participants themself.
Prime examples for this attribution arose during the Cold War era when the Soviet Union was considered to be an enemy by the USA and vice versa.
The last subjective attribution (d) is on the edge of being an objective attribution.
Yet again, the label of "war criminal" is not always objective.
While the definition of a war criminal is codified for example in the Geneva Convention, the allegations of being a war criminal are often disputed.

To verify whether a witness (i.e., a document) from the collection can be seen as a witness we relied on Extractive Question Answering (EQA), i.e., the task of answering a specific question based on a given text \cite{rajpurkar2016squad}.
We formulated a question template for the subjective attributions (a)--(d) masking the  event which can be applied to a witness.\footnote{Example question for (a): "Who was an aggressor in <EVENT\_MASK>"}
In the templates the question was phrased in an open manner to capture synonyms or other representatives of the participants, e.g., statements from nations leaders on behalf of their country.
We used the Hugging Face implementation\footnote{\url{https://huggingface.co/docs/transformers/task_summary\#extractive-question-answering}} in conjunction with a fine-tuned RoBERTa model\footnote{\url{https://huggingface.co/deepset/roberta-base-squad2}} for the EQA task.
Additionally, we defined a threshold for the confidence score of the model to reduce the number of false statements. 

\begin{figure*}
    \centering
    \begin{subfigure}{0.5\linewidth}
        \centering
        \includegraphics[width=0.69\linewidth]{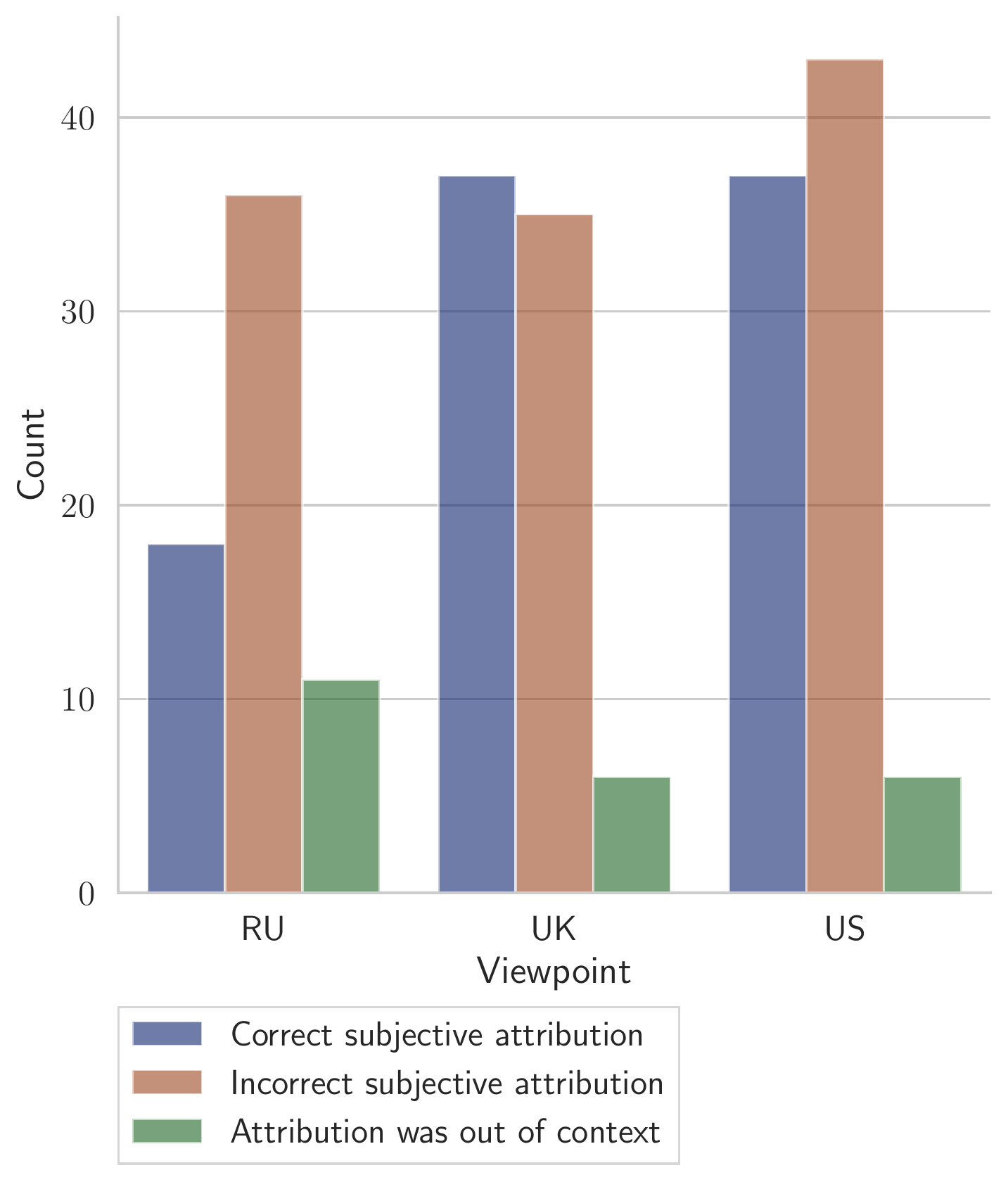}
        \caption{Rater Evaluation Viewpoints}
        \label{subfig:rater_viewpoints}
    \end{subfigure}\hfill
    \begin{subfigure}{0.5\linewidth}
        \centering
        \includegraphics[width=0.69\linewidth]{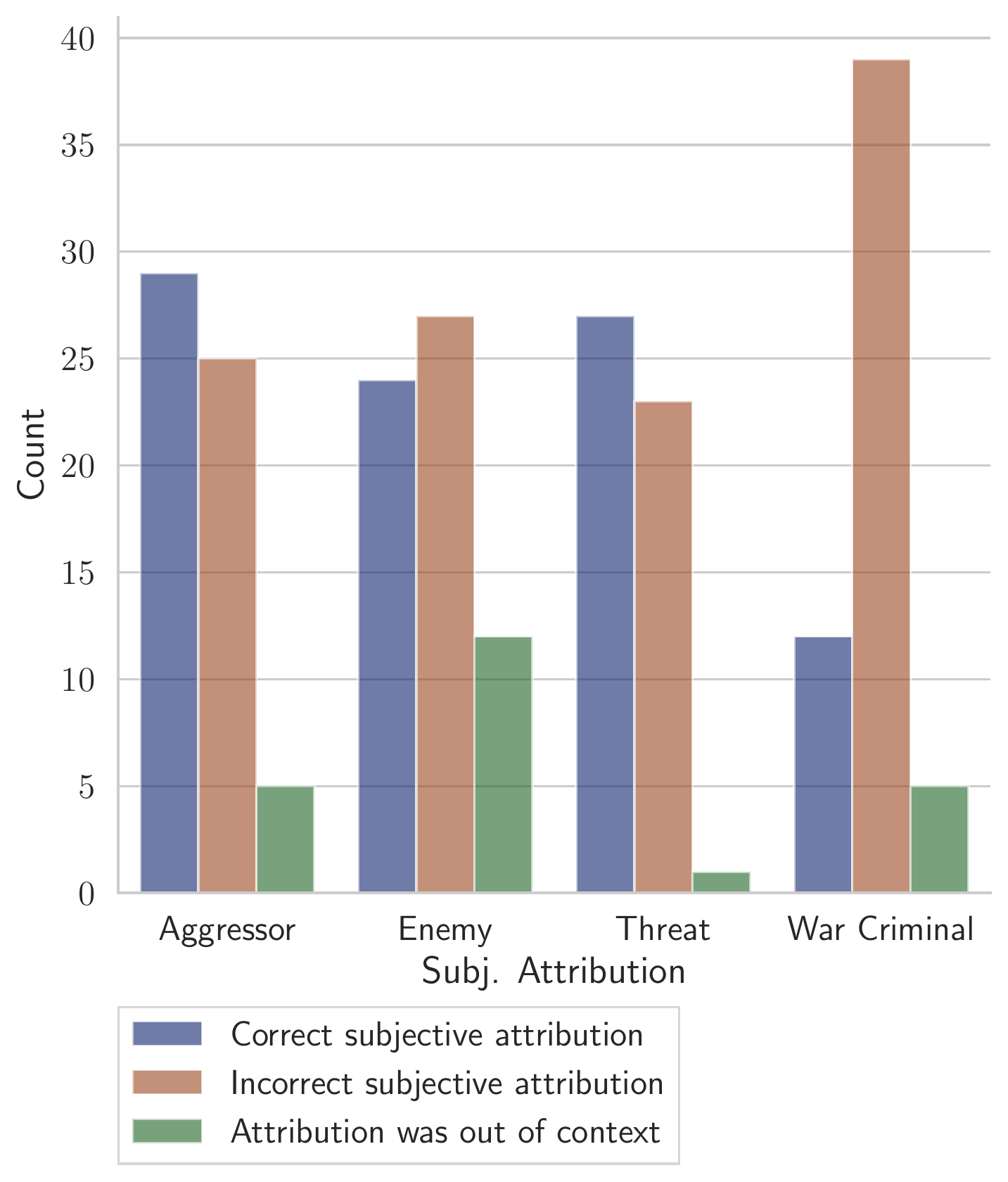}
        \caption{Rater Evaluation Attributions}
        \label{subfig:rater_attributions}
    \end{subfigure}
    \caption{Evaluation of the extractive question answering module.}
    \label{fig:eqa_evaluation}
\end{figure*}
Since the output of EQA is a phrase taken from the witness it has to be canonicalized to a specific participant in order to assess whether the attribution holds for the given participant.
We applied a sentence embedding based on SBERT~\cite{reimers2019sbert} to the EQA output, the attribution candidate, and other participants of the event to adopt the cosine similarity as a measure for canonicalization.
In this process we selected the participant with the highest cosine similarity to the EQA output as attribution target.
If the narrative prototype provided a particular participant for the subjective attribution it was tested, whether the attribution target is identical to the provided participant.

For \textbf{RQ1} the objective is to test whether this approach yields suitable subjective attributions, i.e., does a given witness actually support the subjective attribution.
To test the approach we applied it on the subjective attributions (a)--(d) for each event (1)--(5) and the viewpoints US, UK, and RU with a placeholder as a participant.
We collected the resulting participant along with the top 3 witnesses w.r.t. the EQA confidence score for each view\-point/at\-tri\-bu\-tion/event combination.
We then evaluated manually whether the witnesses state that the attribution is true for the given participant in the event. 
In total we evaluated three witnesses for five conflicts, four subjective attributions, and three viewpoints outlets each, resulting in 180 candidates.

The evaluation was performed by three human raters reaching a fair inter-rater agreement of $\kappa=0.33$ measured by the Fleiss Kappa~\cite{fleiss1971kappa}.
Each rater was provided a set of records where each record encompassed an event, an attribution label (i.e., "aggressor", "war criminal", "threat" and "enemy"), a participant for the attribution, and a link to the witness article.
The task was to categorize each record whether it could be seen as true that the given participant could be attributed in the given way w.r.t. the event, e.g., whether a given article states that "Saddam Hussein" could be seen as a "threat" in the event "Iraq War".
For the task only the headlines and full texts of the articles were used, user comments or other page elements have not been considered by the raters.
To accomplish this task we provided three categories:
\begin{description}
    \item [Correct subjective attribution] The attribution was correct. It was not mandatory that this fact is explicitly stated in the witness document but it should be possible to derive it.
    
    \item [Incorrect subjective attribution] The attribution could not be derived from the witness documents' full text or headline. 
    
    \item[Attribution was out of context] It was possible to derive the attribution but it was taken out of context, e.g., it was part of a quote from another outlet, or otherwise falsely derived.
\end{description}
The results of the evaluation are depicted in Fig.~\ref{fig:eqa_evaluation} rehashed for viewpoints and subjective attributions.
Overall, the incorrect or out of context attributions dominate when using this approach.
Only for the UK viewpoint and the "aggressor" and "threat" attributions, the majority of witnesses seem to support it.
Regardless of the negative outcome, we report on key findings on why the combined approach of EQA and sentence embeddings for canonicalization does not provide sufficient quality for the task.

\emph{Broad attributions and wrong signal words}. The largest outlier in Fig.~\ref{subfig:rater_attributions} is the "war criminal" attribution with mostly incorrect witnesses.
After an inspection of the ratings, these false-positives are mostly based on the second term "criminal" which leads to a lot of out-of-context attributions.
Additionally, sentences like \emph{"Moscow has been throwing its weight around in recent years - in 2008 Russian troops humiliated the Georgians [...]"}~\cite{summers2016dmnato} are a likely cause for an EQA model to mark this sentence as an answer to the question "Who was the war criminal" nonetheless.
The key term here is "humiliation", even though for a human reader this sentence does not imply any war crimes committed by Russia and the remainder of the article does not make any statements in this regard.
The problem of wrong signal words can also be observed for instance for "aggressor" attributions.
Some articles use the term "Iraq invasion" to describe the Iraq War.
However, this term could also be interpreted as a signal for aggressive behavior from Iraq by the EQA model.
Finally, the same problem occurs in enumerations such as \emph{"From the Gulf of Tonkin Resolution to the Mai Lai massacre, the bombing of North Vietnam, [...]"}~\cite{gonchar2017vietnamwar} which ended to an attribution witness to North Vietnam as aggressor although North Vietnam was not mentioned again in this article.

\emph{Problem of negations and context}. Another problem identified during the evaluation concerns out-of-context attributions and missed negations of statements.
Specifically a number of articles was taken as witness for the opposite claim made in the article.
As an example, in an RT.com article~\cite{camp2019liedintowar} regarding claims that the Iraq war was based on lies of Iraq being somewhat close to a "war criminal" and "aggressor".
Nevertheless, the article was considered to be a witness for both attributions beside claiming the exact opposite.
Another example for this behaviour can be observed in an article~\cite{sputniknews2022belarus} about a joint force between Belarus and Russia in the current Russian-Ukrainian tensions which attributed Russia to be an "enemy" beside being heavily framing the EU and NATO as an enemy.
A related issues concerns the context of attributions in articles.
On occasion articles cite other sources or take a quote from persons to take a stance on those quotes.
This, however, occasionally lead to a high EQA score for this articles because the actual quote was taken as witness beside the articles' general stance.

\emph{Corpus related problems}. The last class of problems concerns the dataset itself.
While a brief quality check mostly filtered out articles missing the main topic, a larger number of articles overlapped.
This is especially the case for the Russian-Ukrainian conflict and the Crimea crisis.
The DDG search strings included dates but apparently more time must be spent in efficient pre-processing.

\subsection{Indexing Subjective Attributions}
\label{subsec:indexing}
Research question \textbf{RQ2} concerns suitable index structures for the evaluation of narrative prototypes.
As explained in Sec.~\ref{subsec:np_evaluation} we used a two-tiered index for viewpoints and subjective attributions.
For each viewpoint, participant, and subjective attribution we indexed, whether the subjective attribution holds for the participant for any event with witnesses for the particular viewpoint.
We utilized a Bloom Filter \cite{bloom1970filter} for each viewpoint/attribution combination.
The reason for choosing the Bloom Filter is, that it guarantees the absence of false negatives.
The worst-case scenario here would be a false positive which would only impact the run-time of a query due to an unnecessary evaluation of a subjective attribution.
A false negative would directly impact the expressiveness of the query since a potential match could be omitted.

To test the index we created four narrative prototypes:
\begin{description}
    \item [DvG] This prototype is based on example (a) in Fig.~\ref{subfig:npt_ex1}.
    The prototype matches any event with a supertype \emph{conflict} and an is\_underdog attribution for the participant in event role \emph{winner}.
    \item [DvG+A] In addition to DvG this prototype adds is\_aggressor subjective attribution for at least one participant in the event regarding at least one viewpoint.
    \item [RvU] Example (b) in Fig.~\ref{subfig:npt_ex2} is utilized here. 
    It matches for the current Russian-Ukrainian conflict if is\_aggressor for the US and UK viewpoint evaluates to true.
    \item [CP] The last example prototype matches for every conflict event if is\_aggressor holds  for the United States according to the RU viewpoint and is\_enemy holds  for Russia from the US viewpoint.
\end{description}

\begin{figure}
    \centering
    \includegraphics[width=0.98\linewidth]{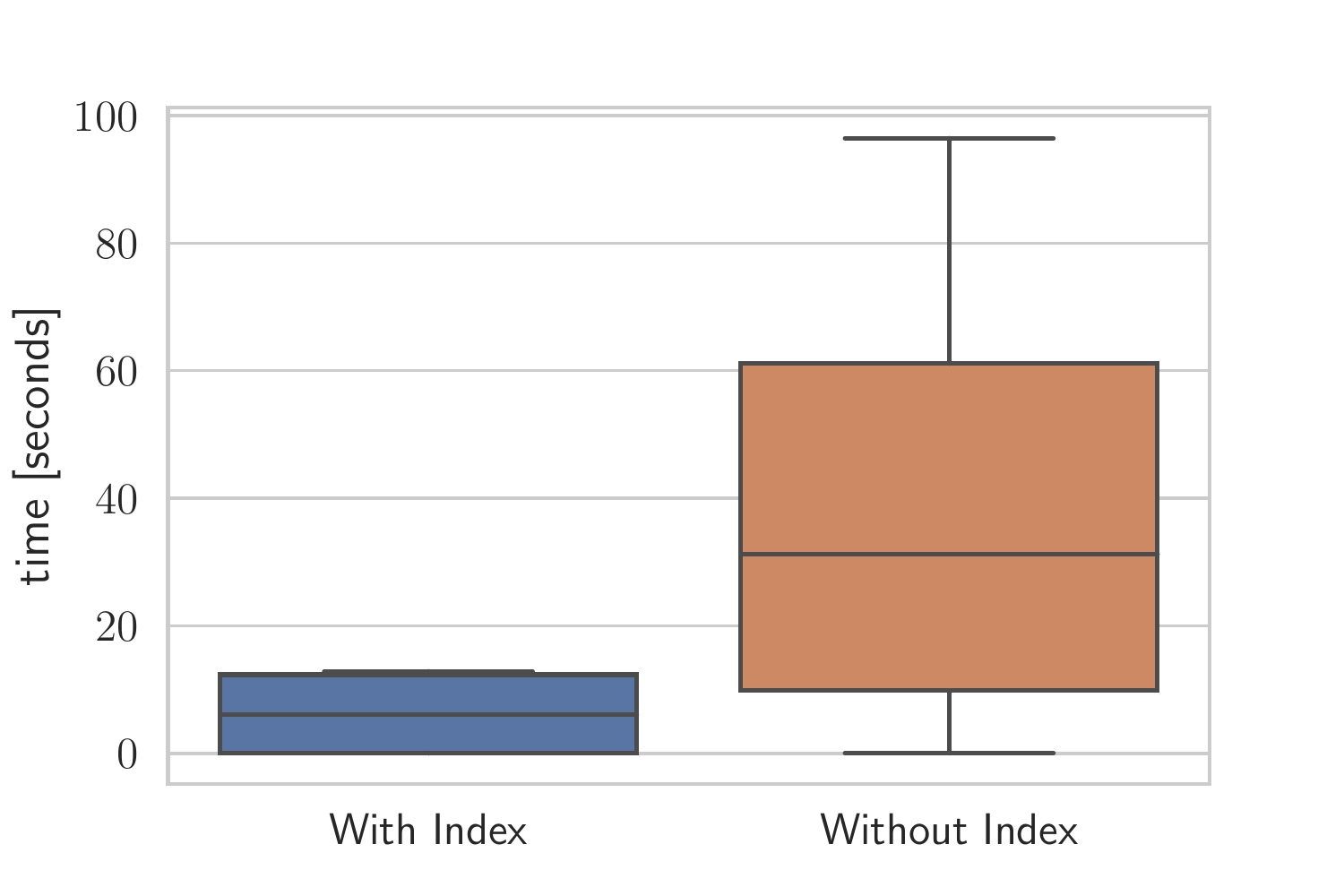}
    \caption{Evaluation Time Comparison}
    \label{fig:time_comparison}
\end{figure}

We collected the run-times of each prototype in two runs using the tailored index in one of the runs.
Index lookups have been performed for each subjective attribution and each possible participant.
If a participant is not located in the corresponding index for a subjective attribution, the evaluation of the attribution is omitted for the respective participant.
The results are depicted in Fig.~\ref{fig:time_comparison}

Without the index, the queries took 3 times as long.
The lower whiskers in both plots are  \textbf{DvG}  which does not rely on subjective attributions, the upper whisker is \textbf{CP} for the non-indexed run and \textbf{RvU} for the indexed run. 
For the non-indexed, run \textbf{CP} took the longest execution time because it required the evaluation of two subjective attributions.
\textbf{RvU} was slightly slower in the evaluation for the indexed run than \textbf{DvG+A}. In both cases, the index reduced the run time.
In general, a higher number of subjective attributions or broader subjective attributions which only require one arbitrary participant are costly to evaluate.
Thus it is essential to reduce the number of participants before the subjective attributions are evaluated.
In this regard, the indexes can help to filter out all participants for which the respective subjective attribution will not be evaluated to true.
Again, due to the usage of Bloom filters, false negatives can not occur and the query evaluation will not lose expressiveness (because in no case the evaluation of a subjective attribution is erroneously omitted) but it increases its effectiveness (because each omitted evaluation of a subjective attribution reduced the total query time).
In summary, index-based queries will always be faster or at least as fast as non-indexed queries without losing expressiveness. 

%% file: sections/05-RelatedWork.tex
In most parts this work relates to the fields of querying subjective data, event representation, and enrichment, narrative structures in information systems, and in part viewpoint detection.

\emph{Subjective Queries}. Research in the direction of querying subjective data has recently gained attention mostly for experiential search for relational data\-bases \cite{li2019subjectivedbs, evensen2019voyager} and conversational search services \cite{gaci2021subjectiveconversation}.
For the most part those systems and methodologies utilize user reviews from platforms like Booking.com or yelp.com and mostly focuses on incorporating opinions from those sources.
Our work also relies on subjective aspects and is similar to \cite{li2019subjectivedbs} since both works bridge structured repositories with subjective data from textual sources.
However, our main focus lies in the enrichment of event-centric knowledge graphs for queries by narrative prototype in contrast to enriching relational databases with subjective attributes to be used in SQL queries.

\emph{Event-Centric Knowledge Graph Enrichment}. The problem of representing events for structured repositories like KGs have been tackled multiple times over the last two decades (see \cite{shaw2009lode, scherp2009feventmodel, van2011design} and in parts \cite{spitz2016termsoverload}).
In this paper we only denoted some basic requirements a representation has to fulfill to be compliant with our query processor. 
As already discussed in Sec.~\ref{subsec:ev_representation_kgs}, we rely on a knowledge graph enrichment instead of representing subjective data directly in the graph.
Our work is most similar to recently published works regarding the enrichment of knowledge graphs with news articles \cite{rospocher2016kgnews, rudnik2019searching} although our approach is not centered around news articles.
In general all kinds of documents suitable to be a witness for subjective attributions are possible enrichments for our system.
Additionally, while \cite{rudnik2019searching} allows for querying the news corpora with a faceted search, querying subjective attributes are not possible.

\emph{Narrative Structures for Information Systems}. Modeling and utilizing narrative aspects for a variety of purposes have gained attention in recent years.
Most works in this direction fall into one of two categories: 
First, the unsupervised extraction of narrative patterns, i.e., script learning \cite{pichotta2014statscriptlearning} or inferring event chains \cite{chambers2008narrativeeventchains, chambers2009narrativeschemas}.
Secondly, the top-down approach, i.e., modeling narrative structures and verify them against a (heterogeneous) knowledge repository \cite{kroll2020narrativestructures, kroll2021narrativequerygraphs}.
Our work falls into the second category.
In contrast to former works we do not focus on a narrative structure in terms of storylines but only on narrative aspects, i.e., attributions for events and participants in those structures.
Additionally, former works did not account for different viewpoints in the modeling process.

\emph{Viewpoint Detection}. Research regarding the detection of viewpoints \cite{thonet2017viewpointdiscovery, quraishi2018viewpoint} and stances (see \cite{aldayel2021stancedetection} for a recent survey in this field) have become increasingly popular in the last years.
While those works mostly focus on the detection of viewpoints (clusters of similar opinions) and stances on certain topics in texts our work utilizes viewpoints only to define which documents can be used as witnesses for subjective attributions.
Future works may include fine-grained viewpoints based on extracted viewpoints from the corpus.\balance

%% file: sections/06-Conclusion.tex
In this paper we introduced \emph{narrative query processing} in conjunction with \emph{narrative prototypes} as a new way to query for narrative aspects in event-centric KGs.
We defined objective and subjective attributions and designed a system for evaluating narrative prototypes based on those concepts.
In a proof of concept we constructed a narrative query processor for both kinds of attributions featuring a number of societal important events and related news articles.
Our experiments focused on the evaluation of subjective prototypes concerning expressiveness and efficiency of narrative queries.

Regarding the query expressiveness, we developed a hybrid query processor that combines structural knowledge bases and unstructured information from texts.
As our experiments showed, narrative prototypes are an effective means of retrieving event-centric information.
Weaknesses in the process were observed during the witness assessment.
Here, efficiently pre-processed corpora can highly improve the assessment quality.
Furthermore, advances in NLP, especially in question answering tasks, will improve the process.
As for efficiency we developed specifically tailored indexes based on Bloom filters.
Due to the design of Bloom filters the expressiveness of narrative queries was preserved but the efficiency was strongly improved.
The main factor here was the reduced number of participants to evaluate for subjective attributions.
For our proof of concept narrative queries were improved by a factor of 3 for a set of narrative queries.
Our indexing schema therefore enables efficient query processing that can be evaluated on a larger scale in the future.
All in all the research in this paper offers a new paradigm to enrich event-centric knowledge graphs by narrative aspects for purposes of expressive querying.

However, the work is limited in two ways: First, the proof of concept is based on a small scale experiment. Due to the novelty of the task no benchmarks are available yet and additional work has to be done to solidify the initial results on a large scale. Secondly the task of finding subjective attributions itself needs to be sharpened. As explained in section \ref{subsec:witness_assessment}, some subjective attributions are overlapping and lack a specific semantic. 
Therefore, future work will be done to mitigate both limitations. 
In particular, the attribution semantics, the quality of the corpus, and the retrieval techniques to determine potential witnesses will be improved.
Additionally, future research may include different levels of granularity for viewpoints by utilizing viewpoint and stance detection techniques in addition to the corpora-level viewpoints.   
Another aspect are questions regarding the weighting of witnesses depending on  users' preferences.